\documentclass[aps,preprint,a4paper,unsortedaddress,onecolumn]{revtex4}

\usepackage[fleqn]{amsmath}
\usepackage{amssymb}
\usepackage[dvips]{graphicx}
\usepackage{color}
\usepackage{tabularx}
\usepackage{mathtools}
\usepackage{algpseudocode}
\usepackage{enumitem}


\newcommand{\vect}[1]{\textbf{\textit{#1}}}

\newcommand{\dir}{\mathrm{dir}}
\newcommand{\rec}{\mathrm{rec}}
\newcommand{\corr}{\mathrm{correction}}
\newcommand{\correlation}{\mathrm{corr}}
\newcommand{\erfc}{\mathrm{erfc}}

\newcommand{\homo}{\mathrm{homo}}

\newcommand{\sinc}{\mathrm{sinc}}

\newcommand{\qh}{q_{\mathrm{H}}}
\newcommand{\qo}{q_{\mathrm{O}}}
\newcommand{\sh}{\vect s_{\mathrm{H}}}
\newcommand{\so}{\vect s_{\mathrm{O}}}
\newcommand{\myphi}{\varphi}
\newcommand{\myphic}{\varphi_c}
\newcommand{\hmyphi}{\hat{\varphi}}
\newcommand{\hmyphic}{\hat{\varphi}_c}
\newcommand{\wshape}{h}

\newcommand{\mo}{\mathcal {O}}
\newcommand{\me}{\mathcal {E}}

\newcommand {\newparagraph} {\vskip .3cm\noindent}

\begin{document}

\title{  
  Kaiser-Bessel Basis for the Particle-Mesh Interpolation }
\author{Xingyu Gao}
\affiliation{Laboratory of Computational Physics, Huayuan Road 6, Beijing 100088, P.R.~China}
\affiliation{Institute of Applied Physics and Computational Mathematics, Fenghao East Road 2, Beijing 100094, P.R.~China}
\affiliation{CAEP Software Center for High Performance Numerical Simulation, Huayuan Road 6, Beijing 100088, P.R.~China}
\author{Jun Fang}
\affiliation{Institute of Applied Physics and Computational Mathematics, Fenghao East Road 2, Beijing 100094, P.R.~China}
\affiliation{CAEP Software Center for High Performance Numerical Simulation, Huayuan Road 6, Beijing 100088, P.R.~China}
\author{Han Wang}
\email{wang_han@iapcm.ac.cn}
\affiliation{Institute of Applied Physics and Computational Mathematics, Fenghao East Road 2, Beijing 100094, P.R.~China}
\affiliation{CAEP Software Center for High Performance Numerical Simulation, Huayuan Road 6, Beijing 100088, P.R.~China}

\newpage
\begin{abstract}
  In this work, we introduce the Kaiser-Bessel interpolation
  basis for the particle-mesh interpolation
  in the fast Ewald method.
  A reliable \emph{a priori} error estimate
  is developed to measure the accuracy of the force computation {in correlated charge systems}, 
  and is shown to be effective in
  optimizing the shape parameter of the Kaiser-Bessel basis in terms of accuracy.
  By comparing the optimized Kaiser-Bessel basis with the
  traditional B-spline basis, 
  we demonstrate that the former is 
  more accurate than the latter in part of the working parameter space,
  saying
  a relatively small real space cutoff,
  a relatively small reciprocal space mesh and a relatively large
  truncation of basis.  
  In some cases, the Kaiser-Bessel basis is found to be more than one order
  of magnitude more accurate.
\end{abstract}

\maketitle

\section{Introduction}

The accurate and efficient computation of the electrostatic
interactions is of central importance in the molecular dynamics
simulations that use the point-charge all-atom force fields~\cite{cisneros2013classical}.
One important type of computational methods is the cutoff method
that approximates the long-range electrostatic interaction
by a short-range spherically cutoff interaction, 
and some examples are
the reaction field method~\cite{onsager1936electric,vangunsteren1978inclusion},
the Wolf method~\cite{wolf1999exact},
the isotropic periodic sum method~\cite{wu2005isotropic} and
the zero-multipole
method~\cite{fukuda2011molecular,fukuda2013zero,wang2016critical}.
Another type, which fully treats the long-range part
of the electrostatic interaction
under the periodic boundary condition,
is the Ewald summation~\cite{ewald1921die} and
its derived fast methods, for instance,
the particle mesh Ewald (PME) method~\cite{darden1993pme},
the smooth particle mesh Ewald (SPME) method~\cite{essmann1995spm},
the particle-particle-particle-mesh (PPPM) method~\cite{deserno1998mue1}
and the multiple staggered mesh Ewald (MSME) method~\cite{cerutti2009multi,wang2016multiple}.
{Ref.~\cite{arnold2013comparison} provided a thorough analysis on the performance of
  the fast Ewald methods and other methods that also treat the long-range part
  of the electrostatic interaction in periodic boundary systems.
}

\newparagraph
{The essential ideas of the fast Ewald methods are the same~\cite{ballenegger2012convert}, 
that is} to
interpolate the charged particles
by a charge distribution defined on a uniform mesh (the particle-mesh interpolation),
and then the computation of the structure factor can be accelerated by
the fast Fourier transform (FFT) that computes the discrete Fourier
transforms with substantially lower complexity.
The accuracy of the particle-mesh interpolation largely relies on the interpolation basis.
The PME method uses Lagrange basis, while the SPME, PPPM and MSME methods use
the B-spline interpolation basis, 
and the later has been shown to be more accurate~\cite{deserno1998mue1}.


\newparagraph
The computation of the structure factor can be viewed
as a nonequispaced discrete Fourier transform (NDFT) of the particle charges.
The fast algorithm of NDFT,
i.e.~the nonequispaced fast Fourier transform (NFFT),
shares almost the same idea
as the fast Ewald method,
and has been developed for a long time
by the applied mathematics community~\cite{dutt1993fast,steidl1998note,ware1998fast,fourmont2003non,fessler2003nonuniform,greengard2004accelerating,keiner2009using,pippig2013parallel}.
{Using the NFFT with B-spline basis in computing the electrostatic interaction in the particle simulations
  is found in Ref.~\cite{hedman2006ewald,pippig2013parallel,nestler2015fast},
  and the equivalence between the NFFT based method and PPPM method has been established in periodic systems~\cite{arnold2013comparison}.}
In the literature of NFFT, a great amount of effort was devoted to
analyzing the accuracy of NFFT approximation
with different interpolation bases (or the window functions in the words of the NFFT community),
such as the Gaussian, {the B-spline, the Bessel and the Kaiser-Bessel functions~\cite{potts2003fast},
and it is argued that the Bessel and the Kaiser-Bessel functions
are more accurate than the B-spline basis~\cite{jackson1991selection,keiner2009using}.}
A natural question then arises:
How accurate would the fast Ewald method be
when using {the Bessel or Kaiser-Bessel} as the basis for the
particle-mesh interpolation.
{Very recently, this question was partially answered by Nestler~\cite{nestler2016parameter}
  who showed that,
  in a homogeneous and uncorrelated charge system,
  the Bessel
  basis function is, in some cases, more accurate than the B-spline basis
  that is
  the ``golden standard'' of the fast Ewald methods.
}

\newparagraph
In this work, we {propose} the Kaiser-Bessel function as the
interpolation basis in the fast Ewald method.
The main difficulty is to determine the optimal shape
of the Kaiser-Bessel function in terms of best accuracy.
We therefore provide an \emph{a priori} error estimate for the
fast Ewald method that uses the Kaiser-Bessel basis,
{in both uncorrelated and correlated charge systems,}
and
show that
the shape parameter that minimizes the estimated error
is very close to that minimizes the actual error.
{In fact the error estimate established by
  Deserno and Holm~\cite{deserno1998mue2} and that by Nestler~\cite{nestler2016parameter} can be
  extended to the Kaiser-Bessel basis without substantial difficulty, but they only apply in the homogeneous and uncorrelated charge systems.
  Our contribution is to develop the error estimate for the correlated charge systems (like water systems),
  which is more relevant for biomolecular simulations.}

\newparagraph
By using the optimal shape parameter,
the accuracy of using the Kaiser-Bessel basis
is systematically compared with that of using the B-spline basis
{in the TIP3P~\cite{jorgensen1983comparison} water system}.
We show that,
for a smaller direct space cutoff,
a larger truncation of interpolation basis
and a smaller mesh size,
the Kaiser-Bessel may be superior to the B-spline basis.
In some cases,
the Kaiser-Bessel is more than one order of magnitude more accurate.
We also detect the cases that
the Kaiser-Bessel is inferior to the B-spline basis.
Therefore, the choice between the Kaiser-Bessel and the B-spline bases
depends on the settings of  other working parameters of the fast Ewald method.
{We also numerically show that, in the TIP3P water system,
  the Kaiser-Bessel basis is only marginally more accurate than the Bessel basis,
  therefore, both of them can be used as a replacement of the B-Spline basis in certain parameter ranges.}


\newparagraph
This work is organized as follows:
Sec.~\ref{sec:ewald} introduces the Ewald summation.
Sec.~\ref{sec:pme} provides the idea of the fast Ewald method,
which is generalized for any kind of interpolation basis.
Sec.~\ref{sec:err-esti} proposes the error estimate for
the Kaiser-Bessel basis in homogeneous and uncorrelated charge systems.
The reliability of determining the shape parameter by the error
estimate is numerically investigated.
In Sec.~\ref{sec:water}, the error estimate is developed for the correlated charge systems, and is validated in a point-charge water system.
In Sec.~\ref{sec:comp}, the Kaiser-Bessel basis is numerically
compared with the B-spline basis in a broad parameter range.
The parameter region where the Kaiser-Bessel basis is more accurate
is summarized.
The work is concluded in Sec.~\ref{sec:conclusion}.

\section{The Ewald summation}
\label{sec:ewald}

We consider a periodic system with
$N$ point charges $\{q_1, q_2, \cdots, q_N\}$  in the unit cell,
and denote their positions by $\{\vect r_1, \vect r_2, \cdots, \vect r_N\}$.
The energy of the electrostatic interaction between the charges
is subject to the Coulomb's law.
The energy of the unit cell is given by
\begin{align}\label{eqn:ele}
  E = \frac12 \sum_{\vect n}^\ast\sum_{i, j = 1 }^N
  \frac{q_i q_j}{\vert \vect r_{ij} + \vect n\vert},
\end{align}
where $\vect r_{ij} = \vect r_i - \vect r_j$,
and $\vect n = n_1\vect a_1 + n_2\vect a_2 + n_3\vect a_3$
with $(n_1, n_2, n_3)\in \mathbb Z^3$
and $(\vect a_1, \vect a_2, \vect a_3)$ being the unit cell vectors.
When $\vect n = 0$ the inner summation computes the energy of the charges
in the unit cell,
while when $\vect n \neq 0$
the inner summation computes the energy between the unit cell
and the periodic image that is shifted by $\vect n$.
The ``$\ast$'' over the outer summation means that
the inner summation should omit the $i = j$ terms when $\vect n = 0$.
{The source of the factor $1/2$ in front of the summation is
  explained in Appendix A of Ref.~\cite{ballenegger2009simulations}.}

\newparagraph
The Ewald summation splits the Coulomb energy~\eqref{eqn:ele}
into three
parts, the direct part, the reciprocal part and the correction
part, i.e.~$
E = E_{\dir} + E_{\rec} + E_{\corr}$,
with the definitions
\begin {align}\label{eqn:es-dir}
  E_{\dir}
  & =
  \frac12 \sum^{\ast}_{\vect n}\sum_{i,j = 1}^{N}
  \frac{q_iq_j \erfc(\beta \vert\vect{r}_{ij} + \vect{n}\vert)}
  {\vert\vect{r}_{ij} + \vect{n}\vert},
  \\\label{eqn:es-rec}
  E_{\rec}
  & =
  \frac1{2\pi V} \sum_{\vect m \neq 0}
  \frac{\exp(-\pi^2\vect m^2 / \beta^2)}{\vect m^2}
  S(\vect m) S(-\vect m), \\\label{eqn:es-cor}
  E_{\corr}
  & =
  -\frac\beta{\sqrt \pi} \sum_{i=1}^N q_i^2,
\end {align}
where $\beta > 0$ is the splitting parameter.
{It is noted here that the surface energy term is absent in
  the Ewald summation due to the spherical summation order in Eq.~\eqref{eqn:ele}
  and the metallic boundary condition.
  The detailed investigation on the surface energy term is found
  in Ref.~\cite{ballenegger2009simulations,de1980simulation,de1980simulation2}. } 
In Eq.~\eqref{eqn:es-rec}, $\vect m = m_1\vect a_1^\ast + m_2\vect a_2^\ast + m_3\vect a_3^\ast$
with  $(m_1, m_2, m_3)\in \mathbb Z^3$ and
$(\vect a_1^\ast, \vect a_2^\ast, \vect a_3^\ast)$ being the reciprocal unit cell vectors
that are defined by
$\vect a_\alpha\cdot \vect a_\gamma^\ast = \delta_{\alpha\gamma}$, $\alpha,\gamma = 1,2,3$.
$V = \vect a_1 \cdot(\vect a_2\times\vect a_3)$
is the volume of the unit cell.
The $\erfc (x)$
in Eq.~\eqref{eqn:es-dir} is the complementary error function, and 
the $S(\vect m)$ in Eq.~\eqref{eqn:es-rec} is the structure factor defined by
\begin{align}\label{eqn:sf}
  S(\vect m) = \sum_{j=1}^N q_j e^{2\pi i \vect m\cdot \vect r_j}.
\end{align}
The ``$i$'' at the exponent is the imaginary unit
that should not be confused with the particle index $i$.

\newparagraph
The complementary error function converges to zero exponentially fast as
the distance between atoms  increases,
therefore, the summation in the direct energy~\eqref{eqn:es-dir}
can take into account a finite number of ``neighbors'' of atom $i$.
More precisely, if the distance between atoms $i$ and $j$
(that is $\vert \vect r_{ij} + \vect n\vert $,
where the periodic images of $j$ should be considered)
is larger than a cutoff radius $r_c$, then the direct interaction between
them can be neglected.
This approximation introduces error in the direct part computation,
but it allows the usage of
the neighbor list algorithm~\cite{frenkel2001understanding}
that reduces the computational complexity to $\mo(N)$.
The summation in the reciprocal energy~\eqref{eqn:es-rec} converges
exponentially fast as one includes more terms.
Therefore, the infinite summation can be truncated as
$-K_\alpha/2 \leq m_\alpha < K_\alpha/2$,
and the number of summed terms should take~$K_1 K_2 K_3 \sim N $,
which leads to an overall computational complexity of $\mo(N^2)$.
When using a larger splitting parameter $\beta$,
the complementary error function in the direct energy~\eqref{eqn:es-dir} converges faster,
so the energy contribution due to the neighbors outside the cutoff
radius becomes smaller,
and the accuracy of the direct space computation increases.
At the same time,
the exponential in the reciprocal energy~\eqref{eqn:es-rec} converges slower,
so the accuracy of the reciprocal space computation decreases.
Therefore, 
a smaller cutoff $r_c$ and a larger truncation $K_\alpha$
should be used with a larger $\beta$,
which leads to a decreased computational cost of the direct part and
an increased computational cost of the reciprocal part, respectively.
Thus, the splitting parameter provides a way to move the
computational load from the direct to the reciprocal part
of the Ewald summation and \emph{vise versa}.
It has been pointed out that the optimal computational cost of
$\mo(N^{1.5})$ is achieved when the direct and reciprocal computational
costs are balanced~\cite{perram1988asc}.

\section{The fast Ewald method using the Kaiser-Bessel basis }
\label{sec:pme}

The computational cost of $\mo(N^{1.5})$ becomes too expensive
for simulations of systems that have more than a thousand of atoms.
The development of the fast Ewald methods
starts from an observation:
If the atoms in the system  locate on a $K_1 \times K_2 \times K_3$
uniform mesh and $N = K_1 K_2 K_3$,
then the structure factor is nothing but a discrete Fourier transform
of the charge distribution that can be computed at the cost of
$\mo (N \log N)$ by using the FFT.
In practice, as the atoms are usually not on the uniform mesh,
the charge distribution is firstly \emph{interpolated}
on the uniform mesh,
and then the computation on the mesh is accelerated by the FFT.
Finally, the resulting force, defined on the mesh,
is interpolated back to the particles.
The computational cost of the interpolation grows in proportion to
the number of particles,
and that of the FFT grows as $\mo(N\log N)$,
thus the total computational complexity in the reciprocal space is $\mo(N\log N)$.
Since the direct space cost is $\mo(N)$,
the overall computational complexity of the fast Ewald methods
is $\mo(N \log N)$.

\newparagraph
In the following, we briefly introduce the fast Ewald method.
For more details, the readers are referred to the literature like Refs.~\cite{darden1993pme,essmann1995spm,deserno1998mue1}.
We let the number of mesh points on direction $\alpha$ ($\alpha = 1, 2, 3$) be
$K_\alpha$, and rescale the $\alpha$ component of the particle coordinate $\vect r$
to $u_\alpha$ 
by 
$
u_\alpha = K_\alpha r_\alpha
$, {where  $r_\alpha =  \vect a_\alpha^\ast \cdot \vect r$.}
The range of $u_\alpha$ is $[0, K_\alpha)$.
By using this notation, the interpolation of the single particle contribution to the
structure factor,
$q e^{2\pi i m_\alpha u_\alpha/K_\alpha}$,
is given by 
\begin{align}\label{eqn:approx-exp-1d}
  qe^{2\pi i m_\alpha u_\alpha/K_\alpha} \
  \approx
  qf(u_\alpha) :=  
  \frac{\hat g(m_\alpha) }{K_\alpha}
  \sum_{l_\alpha \in I_{K_\alpha}} q\,\myphic (u_\alpha -l_\alpha)
  \,e^{2\pi i m_\alpha l_\alpha/K_\alpha},
\end{align}
where  $I_{K_\alpha} = \{ \: l \in \mathbb Z \:\vert -K_\alpha/2 \leq l < K_\alpha/2\}$.
The interpolation basis is denoted by $\myphi$, and $\myphic$ is the truncated interpolation basis
that is defined by
\begin{align}
  \myphic (x_\alpha) =
  \begin{dcases}
    \myphi(x_\alpha) & \vert x_\alpha \vert \leq C, \\
    0 & \mathrm{otherwise\ in\ } I_{K_\alpha},
  \end{dcases}
\end{align}
with the truncation radius $C$ being a positive integer.
{Outside interval $I_K$, $\myphic$ is periodically extended with period $K_\alpha$ to $\mathbb R$.}
With the truncation, the interpolation of one charge~\eqref{eqn:approx-exp-1d}
can be computed at a constant computational cost, and the cost of
interpolating all charges in the system is $\mo(N)$,
otherwise, the interpolation would take $N K_1K_2K_3\sim N^2$ operations.
Therefore, the truncation is always enforced, and 
it is usually required that the interpolation basis converges to zero fast
outside the truncation radius.
In Eq.~\eqref{eqn:approx-exp-1d},
$\hat g(m_\alpha)$ is a prefactor that can be chosen in several ways.
For example, the SPME method takes~\cite{essmann1995spm,ballenegger2012convert}
\begin{align}\label{eqn:b-spme}
  \hat g(m_\alpha) = 
  \frac{1}{\sum_{l_\alpha} \hmyphic(m_\alpha + l_\alpha K_\alpha)},
\end{align}
while the PPPM method that optimizes the reciprocal force accuracy 
in a homogeneous and uncorrelated charge system derives
\begin{align}\label{eqn:b-pppm}
  \hat g(m_\alpha) = 
  \frac
  {\hmyphic(m_\alpha)}
  {\sum_{l_\alpha} \hmyphic^2(m_\alpha + l_\alpha K_\alpha)},
\end{align}
where 
the $\hmyphic$ denotes the Fourier transform of truncated interpolation basis.
In this work, we adopt the PPPM convention~\eqref{eqn:b-pppm}, 
and explain this expression in Appendix~\ref{app:a}.

\newparagraph
Inserting Eq.~\eqref{eqn:approx-exp-1d} in the reciprocal part of Ewald summation \eqref{eqn:es-rec},
we arrive at
\begin{align}\label{eqn:approx-rec-ener}
  E_\rec
  \approx \ &  
  \sum_{l_1,l_2,l_3}
  Q(l_1,l_2,l_3)
  [\,Q \ast (F B^2)^{\vee}] (l_1,l_2,l_3),
\end{align}
where
\begin{align} \label{eqn:b}
  &B(\vect m) = \prod_{\alpha} \hat g(m_\alpha), \\ \label{eqn:f}
  &F(\vect m)
  =
  \frac1{2\pi V}\times
    \begin{dcases}
      \frac{\exp(-\pi^2\vect m^2 / \beta^2)}{\vect m^2} &  \vert \vect m \vert \neq 0, \\ 
      \: 0 & \vert \vect m\vert  = 0,
    \end{dcases} \\\label{eqn:p}
  &P_{\vect r}(l_1,l_2,l_3) = \prod_\alpha \myphic(u_\alpha - l_\alpha), \\ \label{eqn:q}
  &Q (l_1,l_2,l_3)= \sum_j q_j P_{\vect r_j} (l_1,l_2,l_3).
\end{align}
The symbol ``$\ast$'' denotes the convolution, and ``$\vee$'' denotes
the backward discrete Fourier transform.
The computational complexity of Eq.~\eqref{eqn:q}
is $\mo(N)$, because the interpolation basis $\myphi$ is truncated.
The convolution $Q \ast (F B^2)^{\vee}$
in Eq.~\eqref{eqn:approx-rec-ener}
can be computed at a cost of $\mo(N \log N)$ by using the
identity $Q \ast (F B^2)^{\vee} = [\, \hat Q \times (F B^2)\, ]^\vee$
and the fast Fourier transform.

\newparagraph
The reciprocal force of particle $i$ is similarly approximated by
\begin{align}\label{eqn:ik}
  \vect F_{\rec,i}
  \approx \ &
  q_i  
  \sum_{l_1,l_2,l_3}
  P_{\vect r_i} (l_1,l_2,l_3)
  [\,Q \ast (\vect G B^2)^{\vee}] (l_1,l_2,l_3),
\end{align}
with notation
\begin{align}\label{eqn:G}
\vect G(\vect m) = -4\pi i\vect m F(\vect m).
\end{align}
{This force scheme is known as the \textit{ik}-differentiation~\cite{wang2016multiple}.
  The other popular scheme, i.e.~the analytical differentiation,
  which takes the differentiation on the approximated energy Eq.~\eqref{eqn:approx-rec-ener} with respect to the particle position,
  cannot be constructed if the interpolation basis $\myphi$
  is not differentiable.
  The Kaiser-Bessel basis investigated in this work
  is discontinuous at the truncation $C$, thus using the analytical
  differentiation to compute the particle force is impossible for this basis.
}

\newparagraph
In Eq.~\eqref{eqn:approx-exp-1d}, one has the freedom of choosing the interpolation basis $\myphi$.
The original PME method uses the Lagrange interpolation, while
the SPME and PPPM methods use the cardinal B-spline interpolation, which is shown to be better than Lagrange interpolation in terms of accuracy~\cite{deserno1998mue1}.
The $n$-th order B-spline basis can be defined in a recursive way:
\begin{align}
  \varphi_{1} (x) &= \chi_{[-\frac12,\frac12]}(x), \quad
  \myphi_n(x) = K \varphi_{n-1} \ast \varphi_{1} (x),
\end{align}
where $\chi_{[-1/2,1/2]}$ denotes the characteristic function on interval $[-1/2, 1/2]$.
$\hmyphi(m )$, the Fourier transform of $\myphi$, is given by
\begin{align}\label{eqn:hat-phi}
  \hmyphi_n(m )  =  \frac1{K} \Big[\sinc( \frac{\pi m }{K} )\Big]^n.
\end{align}
It should be noted that the $n$-th order B-spline basis has
a compact support of $[-n/2,n/2]$, therefore,
it is natural to take the truncation radius as $C = n/2$, and thus $\myphi \equiv \myphic$.

\newparagraph
{
  The Bessel basis is defined by
  \begin{align}
    \myphi(x) = I_0 (\pi h \sqrt{C^2 - x^2}),
  \end{align}
  where $h$ is the shape parameter, and $I_0$ denotes the modified {zero-order} Bessel function. Its Fourier transform is
  \begin{align}
    \hmyphi(m) =
    \frac{1}{K} \times
    \begin{dcases}
      \frac{\sinh (\pi C \sqrt{h^2 - (2m/K)^2 } ) }{\pi\sqrt{h^2 - (2m/K)^2 }}    
      &\quad \big\vert {2 m}/{K} \big\vert \leq h, \\
      \frac{\sin (\pi C \sqrt{(2m/K)^2 - h^2 } ) }{\pi\sqrt{(2m/K)^2 - h^2 }}
      & \quad \big\vert {2 m}/{K} \big\vert > h,
    \end{dcases}      
  \end{align}
  where $\sinh$ is the hyperbolic sine function defined by $\sinh(x) = (e^x - e^{-x})/2$.
}
\newparagraph
{The Kaiser-Bessel basis, which interchanges the role of the direct and Fourier
domains of the Bessel basis, is defined by}
\begin{align}
  \myphi (x) =
  \frac{\sinh (\pi \wshape \sqrt{C^2 - x^2} ) }{\pi\sqrt{C^2 - x^2}},
\end{align}
where $\wshape$ is  the shape parameter.
Its Fourier transform is given by
\begin{align}\label{eqn:kb-f}
  \hmyphi(m) =
    \frac{1}{K} \times
    \begin{dcases}
      I_0 \Big( \pi C \sqrt{\wshape^2 - \big({2 m}/{K}\big)^2 }\; \Big)
      &\quad \big\vert {2 m}/{K} \big\vert \leq h, \\
      0 & \quad \big\vert {2 m}/{K} \big\vert > h,
    \end{dcases}      
\end{align}
where $I_0$ denotes the modified {zero-order} Bessel function.

\begin{figure}
  \centering
  \includegraphics[width=0.48\textwidth]{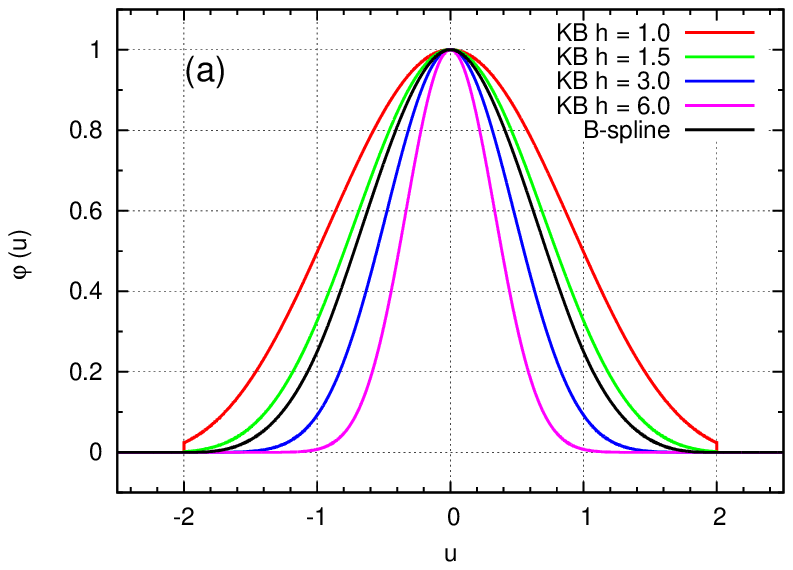}\\
  \includegraphics[width=0.48\textwidth]{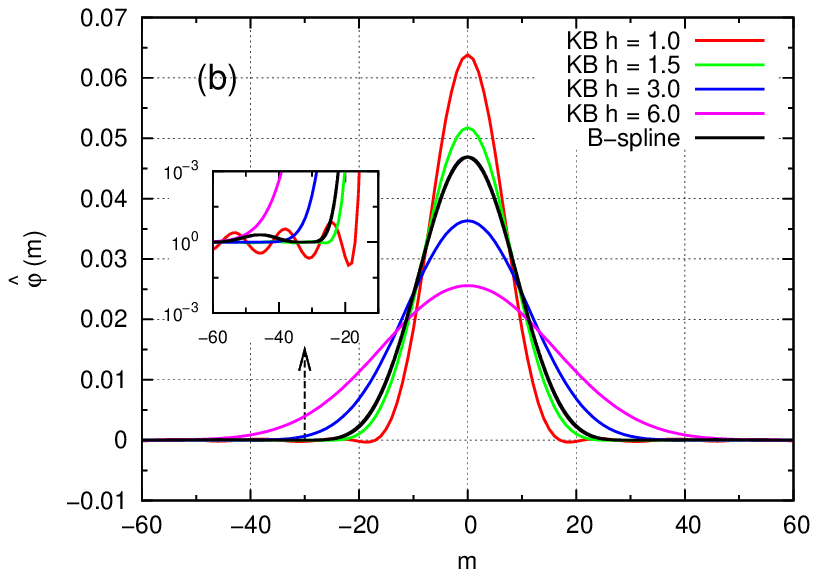}
  \caption{The truncated Kaiser-Bessel basis with different shape parameters
    and the B-spline basis (plot (a)),
    and their Fourier transforms (plot (b)).
    In the legend, the ``KB'' stands for Kaiser-Bessel.
    The truncation radius $C$ is taken to be~2.
    The number of mesh points is $K=32$.
    In the plots, all the basis and the Fourier transform pairs
    are rescaled by the basis value measured at 0 for an easier comparison.
  }
  \label{fig:tmp1}
\end{figure}

\newparagraph
The Fourier transform of the truncated basis $\myphic$,
which is denoted by $\hmyphic$ in this work,
is in general not equal to $\hmyphi$ given by \eqref{eqn:kb-f}, 
thus  it is evaluated numerically.
Although this numerical evaluation
is more expensive than an analytical formula,
fortunately, one only needs to do it once before the simulation starts.
It should be noticed that, unlike the full Kaiser-Bessel basis,
the Fourier transform of the truncated basis
does not have a compact support in the reciprocal space
(see Fig.~\ref{fig:tmp1}).
When using a larger shape parameter $\wshape$,
the Kaiser-Bessel basis is more localized in the direct space,
while its Fourier transform is 
more wide-spread in the reciprocal space (see also Fig.~\ref{fig:tmp1}).

\newparagraph
An important problem is to determine
the shape parameter $\wshape$ for the Kaiser-Bessel basis.
We notice 
the fact that when the truncation radius $C$ is fixed,
the computational cost of particle-mesh interpolation
does not depend on $\wshape$,
so changing $\wshape$ will not alter the computational cost of
the fast Ewald method.
Therefore, we are allowed to tune the shape parameter $\wshape$
in terms of minimizing the reciprocal force error, 
so that the optimal accuracy is obtained from the same computational expense.
In practice, the accuracy of the force computations given a shape parameter
is not easy to know, unless the computed force is compared with
a reference force that is obtained by using exhaustively accurate parameters.
One solution is the error estimate that provides 
an \emph{a priori} and quantitative description of the error
as a function of the shape parameter, 
as well as other working parameters of the fast Ewald method.
Then the shape parameter is determined via minimizing the
estimated error.
In this context, the quality of the error estimate matters,
because the shape parameter will be close to the optimal one
if the error estimate is accurate.
The error estimate and the determination of the shape parameter
will be discussed in Sec.~\ref{sec:err-esti}.


\section{Error estimate and the optimal basis shape}
\label{sec:err-esti}

In order to derive the error estimate,
a clear definition of the word ``error'' is necessary.
In the community of molecular dynamics simulation,
the error is usually defined as the root mean square (RMS)
error in the force computation:
\begin{align}
\me = \sqrt{\langle \vert \Delta \vect F\vert^2 \rangle},  
\end{align}
where the error force $\Delta \vect F$ denotes
the difference  between the computed and the accurate forces,
and the $\langle \cdot \rangle$ denotes the ensemble average.
We refer to the RMS force error as the ``error'' for convenience.
The error is composed by the contributions from both
the direct and reciprocal parts, saying
\begin{align}\label{eqn:err-dir-rec}
  \me^2 = \me^2_\dir + \me^2_\rec,
\end{align}
where $\me_\dir = \sqrt{\langle \vert \Delta \vect F_\dir\vert^2 \rangle}$ and
$\me_\rec = \sqrt{\langle \vert \Delta \vect F_\rec\vert^2 \rangle}$,
with $\Delta\vect F_\dir$ and $\Delta\vect F_\rec$ being the error forces
in the direct and reciprocal spaces, respectively.
The error
is of the additive form~\eqref{eqn:err-dir-rec} because it is reasonable to assume the independency
of the direct and reciprocal error contributions.

\newparagraph
In the direct space, the error originates from neglecting the
particle interactions beyond the cutoff radius,
and has been studied by, for instance, Refs.~\cite{kolafa1992cutoff,wang2012error}.
In the reciprocal space,
the error originates from particle-mesh
interpolation that approximates the particle contribution to the structure factor
by a linear combination of the interpolation bases defined 
on the uniform mesh.
The magnitude of the reciprocal error is 
controlled by the interpolation basis and the mesh size.
For a homogeneous and uncorrelated charge system,
we estimate the reciprocal error of force computation 
by using the error estimate
framework proposed in Ref.~\cite{wang2012numerical}, and reach
\begin{align}\label{eqn:err-esti-homo}
  \vert \me^\rec\vert^2 \approx
  \vert \me^\rec_\homo\vert^2
  =
  2q^2 Q^2
  \sum_{\vect m}
  \vect G^2(\vect m)
  \sum_{\alpha}\sum_l
  Z^2_{\alpha,l}(\vect m)
\end{align}
with
\begin{align}\label{eqn:err-esti-homo-z}
  Z_{\alpha,l}(\vect m)
  =
  \frac{\hmyphic(m_\alpha)\hmyphic (m_\alpha + lK_\alpha)}
  {\sum_l \hmyphic^2(m_\alpha + lK_\alpha) } - \delta _{l0},
\end{align}
where $Q^2 = \sum_i q_i^2$, $q^2 = Q^2 / N$ and $\delta$ is the Kronecker delta.
The magnitude of the error is related to the interpolation basis via
function $Z_{\alpha,l}(\vect m)$.
The outline of the proof is provided in Appendix~\ref{app:a}.


\begin{figure}
  \centering
  \includegraphics[width=0.48\textwidth]{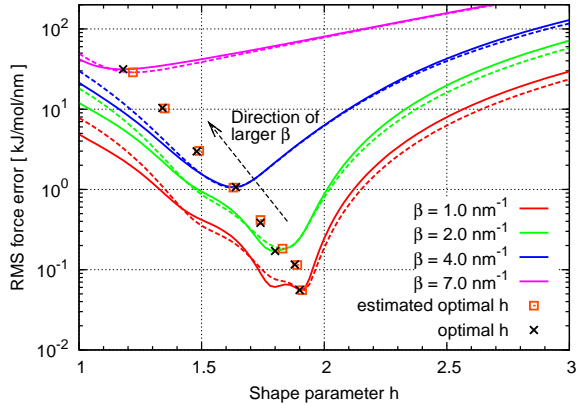}
  \caption{
    The reciprocal error plotted against the shape parameter $h$ of the Kaiser-Bessel interpolation basis.
    The truncation radius of the interpolation basis is $C=2$.
    The number of mesh points on each direction is $K_\alpha = 32$.
    The reciprocal force error is plotted by solid color lines for
    the splitting parameter  
    $\beta = 1.0$~$\mathrm{nm}^{-1}$ (red), 2.0~$\mathrm{nm}^{-1}$ (green),
    4.0~$\mathrm{nm}^{-1}$ (blue) and 7.0~$\mathrm{nm}^{-1}$ (pink).
    The dashed lines are the corresponding error estimates of the solid lines with the same color.
    The black cross indicates the optimal shape parameter that minimizes the reciprocal force error
    for $\beta = 1.0,\ 1.5,\ 2.0,\ 3.0,\ 4.0,\ 5.0,\ 6.0,\ 7.0\ \mathrm{nm}^{-1}$.
    The orange square indicates the optimal shape parameter determined by minimizing the
    estimated error.
  }
  \label{fig:tmp2}
\end{figure}

\newparagraph
As mentioned by
in Sec.~\ref{sec:pme},
the optimal shape of the Kaiser-Bessel basis can be determined by minimizing
the error estimate~\eqref{eqn:err-esti-homo} with respect to the
shape parameter $h$.
We therefore numerically check the quality of the error estimate and its ability
to predict the optimal shape parameter for the Kaiser-Bessel basis
in a uniform and uncorrelated charge system.
The simulation region is of size
$3.724\textrm{nm} \times 3.724\textrm{nm} \times 3.724\textrm{nm}$, 
and contains 5184 randomly distributed charged particles.
One third of the particles (1728) have a negative partial charge  $-0.834\,e$, while the
rest two thirds (3456) have $0.417\,e$.
The whole system is neutral.

\newparagraph
The actual and estimated reciprocal force errors computed in this system are plotted
against the shape parameter $h$ by solid and dashed color lines, respectively, in Fig.~\ref{fig:tmp2}.
The actual error is obtained by comparing the reciprocal force with a well-converged
Ewald reciprocal force computation with the same splitting parameter.
Different colors in the Figure stand for different choices of the splitting parameter $\beta$.
For any $\beta$, the error blows up for either very small or very large $h$,
therefore, there exists an optimal $h$ that minimizes the error,
and an arbitrary choice of the shape parameter $h$ may lead to a computation that
is orders of magnitude less accurate than the optimal one.
The square and cross present the position of the optimal shape
when varying the splitting parameter $\beta$.
The former is determined by minimizing the estimated error, while the latter
is determined by minimizing the actual error.
A clear dependency of the optimal shape on the splitting parameter $\beta$ is shown by the Figure,
thus there does not exist a ``universal'' optimal shape,
and the shape of the basis should be optimized for each distinct splitting parameter.
The \emph{a priori} error estimate provided by us is of satisfactory quality
in the sense that the estimated error is very close to the actual error,
and
the optimal shape parameter $h$ can be closely approximated 
by minimizing the estimated error.


\begin{figure}
  \centering
  \includegraphics[width=0.45\textwidth]{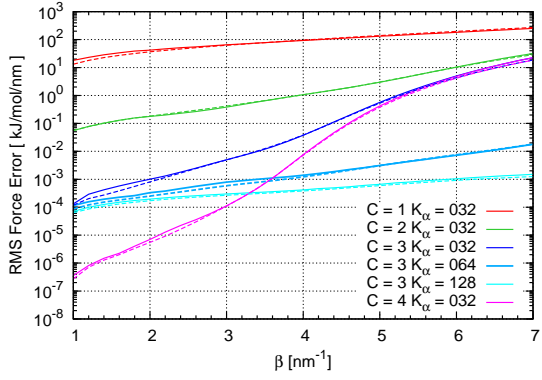}
  \caption{The actual and estimated reciprocal errors of the Kaiser-Bessel basis plotted against the splitting parameter $\beta$ in the homogeneous and uncorrelated charge system.
    The solid lines present the actual error, while the dashed lines present the estimated error.
    Different colors denote different combinations of the truncation radius and the mesh size, which are listed in the legend of the figure.
    The shape parameter is optimized by minimizing the estimated error.
  }
  \label{fig:error-comp}
\end{figure}

\newparagraph
To check the quality of the error estimate in a wider range of parameters,
the estimated reciprocal error is compared with
the actual error for various combinations of the 
reciprocal space working parameters, namely the splitting parameter $\beta$,
the truncation radius $C$ of the basis 
and the number of mesh points  $K_\alpha$ in Fig.~\ref{fig:error-comp}.
For each combination of $\beta$, $C$ and $K_\alpha$,
the shape parameter is optimized by minimizing the estimated reciprocal error.
In the Figure, the actual (solid lines) and estimated (dashed lines)
reciprocal force errors are plotted against the splitting parameter.
The lines with different colors denote different choices of
the basis truncation radius and the mesh size
(see the legend of the figure for all investigated parameters).
All the estimated errors closely follow the actual errors
for all considered parameters,
so the accuracy of using the Kaiser-Bessel basis in the fast Ewald method
can be reliably measured by the error estimate.

\section{Error estimate in the correlated charge systems }
\label{sec:water}

\newparagraph
So far we have developed and tested  the error estimate of Kaiser-Bessel basis in
an uncorrelated charge system.
In practice, the charges in most systems are correlated
for many reasons,
for example, the covalence bonds,
the van der Waals interaction and the hydrogen bonding.
A widely studied  system, in which the charges are correlated,
is water. 
In this work,
we provide the correction to the error estimate due to the
charge correlation of water, which contributes to the reciprocal force error by
\begin{align}
  \vert \me^\rec\vert^2 \approx \vert \me^\rec_\homo\vert^2 + \me^\rec_\correlation.
\end{align}
The correlation error $\me^\rec_\correlation$ can be partially estimated by the nearest neighbor approximation
technique~\cite{wang2012numerical}.
We take the rigid three-point-charge water model
for example~\cite{wang2016multiple}
\begin{align} \nonumber
  \me^\rec_\correlation = &\,
  q^2 Q^2
  \sum_{m}
  \vect G^2(\vect m)T^w(\vect m) 
  \sum_{\alpha,l}
  Z^2_{\alpha,l}(\vect m)  \\
  &+
  q^2 Q^2
  \sum_{m}
  \vect G^2(\vect m)
  \sum_{\alpha,l}  
  T^w(\vect m + lK_\alpha \vect a_\alpha^\ast) Z^2_{\alpha,l}(\vect m) ,
\end{align}
where  $T^w$ is a function that delivers the charge correlation due to the covalence bonds in the water molecule. We have
\begin{align}
  T^w(\vect m)
  & =
  \frac{4\qh \qo}{2\qh^2 + \qo^2} T_{ \so}(\vect m)
  +
  \frac{2\qh^2}{2\qh^2 + \qo^2} T_{\sh}(\vect m),
\end{align}
where $\qo$ and $\qh$ are the partial charges of the oxygen and hydrogen atoms, respectively.
$\so$ is the vector connecting the oxygen and the hydrogen atoms, and
$\sh$ is the vector connecting two hydrogen atoms.
The function $T_{\vect b}(\vect m)$ defined for vector $\vect b$
is the structure factor averaged over all possible directions of $\vect b$
\begin{align}\label{eqn:tb-sin}
  T_{\vect b}(\vect m)
  = \langle e^{2\pi i \vect m\cdot \vect b}\rangle_{\mathrm{directions}}
  =
  \frac{\sin(2\pi m\vert \vect b\vert)}
  {2\pi m \vert\vect b\vert}.
\end{align}
Taking the TIP3P water model for example, $\qo = -0.834$~e, $\qh =
0.471$~e, $\vert \so \vert = 0.09572$~nm and $\vert \sh\vert =
0.15139$~nm.

\begin{figure}
  \centering
  \includegraphics[width=0.45\textwidth]{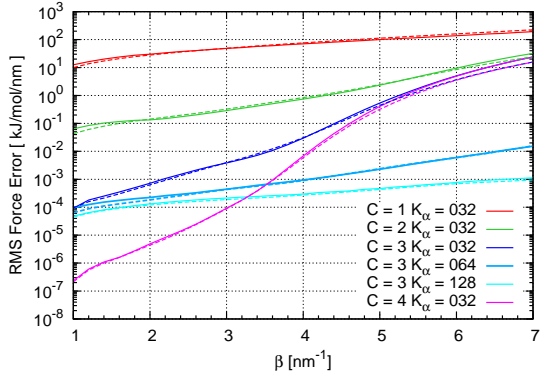}
  \caption{The actual and estimated reciprocal errors of the Kaiser-Bessel basis plotted against the splitting parameter $\beta$ in the TIP3P water system.
    The solid lines present the actual error, while the dashed lines present the estimated error.
    Different colors denote different combinations of the truncation radius and the mesh size, which are listed in the legend of the figure.
    The shape parameter is optimized by minimizing the estimated error.
  }
  \label{fig:error-comp-water}
\end{figure}

\newparagraph
The actual and estimated reciprocal errors of the Kaiser-Bessel basis
in the TIP3P water system are presented in Fig.~\ref{fig:error-comp-water}.
The simulation region is of size
$3.724\textrm{nm} \times 3.724\textrm{nm} \times 3.724\textrm{nm}$, 
and contains 1728 water molecules.
The water configuration is taken from an equilibrium NPT
simulation~\cite{gao2016sampling}.
The charge density in this system is the same as the homogeneous
and uncorrelated charge system introduced in Sec.~\ref{sec:err-esti}.
Similar to Fig.~\ref{fig:error-comp}, the error is plotted
against the splitting parameter with different combinations of the
truncation radius and mesh size
(see the legend for all parameter combinations).
The shape parameter is optimized by minimizing the estimated error.
From the Figure, it is shown that the error estimate is of good quality
so that the actual error is reliably computed by the estimate.

\section{Comparison of the B-spline, Bessel and Kaiser-Bessel bases}
\label{sec:comp}

\begin{figure}
  \centering
  \includegraphics[width=0.40\textwidth]{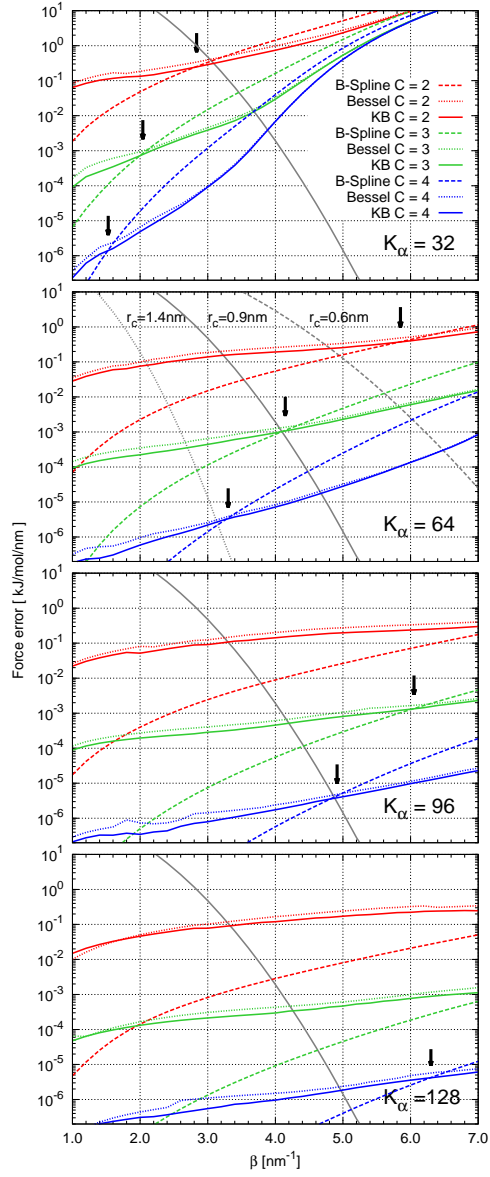}  \\   
  \caption{The comparison between the reciprocal force errors computed by
    {the B-spline basis (dashed lines),
    the Bessel basis (dashed lines) and
    the Kaiser-Bessel basis (KB in short, solid lines)
    in the TIP3P water system.}
    The error as a function of the splitting parameter $\beta$ is presented.
    In each plot, the red, green and blue colors present the errors
    computed with truncation radius $C=2$, 3 and 4, respectively.
    The four plots from up to down take the mesh size of $K_\alpha = 32,\ 64,\ 96$, and 128,
    respectively.
    The crossover of the Kaiser-Bessel and the B-spline bases is
    indicated by the black arrow in the plots.
    The dashed, solid and dotted gray lines present the direct errors
    of cutoff radii 0.6, 0.9 and 1.4 nm, respectively.
  } 
  \label{fig:tmp3}  
\end{figure}

{The reciprocal force errors of
the B-spline, Bessel and Kaiser-Bessel
bases are presented and compared for the TIP3P water system
in Fig.~\ref{fig:tmp3}.
The four plots
in the Figure, from up to down, use $K_\alpha = 32$, 64, 96 and 128
mesh points on each direction, respectively.  In each plot, the errors
of B-spline, Bessel and Kaiser-Bessel bases are plotted {against the splitting parameter $\beta$} by dashed, dotted and solid
lines, respectively, and the truncation radii $C=2$, 3 and 4 are
presented by red, green and blue lines, respectively.
In most cases, the Kaiser-Bessel basis is only marginally more accurate than the Bessel
basis, and there are exceptions ($K_\alpha = 128$, small $\beta$) that  the
Bessel basis is more accurate.
In the range of relatively large splitting parameter $\beta$,
in which the Kaiser-Bessel basis is of particular interest (see the discussion below),
the difference between the two bases is almost negligible.
By contrast, the difference between the B-Spline and the Kaiser-Bessel bases
is more significant, therefore, we focus on comparing them in the rest of this manuscript.
}

\newparagraph
It is observed from the Figure that at relatively large $\beta$,
the Kaiser-Bessel basis is more accurate,
while at relatively small $\beta$,
the B-spline basis is more accurate,
therefore, there exists a crossover of the accuracy of the
two interpolation bases.
When using a splitting parameter larger
than the crossover, the Kaiser-Bessel is more favorable than
the B-spline in terms of accuracy,
and this advantage is observed to be more prominent for the larger truncation radius $C$.
{For example, at} truncation radius $C=4$ and mesh size $K_\alpha = 32$, the Kaiser-Bessel basis can be
more than one order of magnitude more accurate than the B-spline basis (around $\beta = 3.0$~nm).
The crossovers for different parameter settings are indicated by
black arrows in Fig.~\ref{fig:tmp3}.
Two trends of the position of the crossovers can be summarized:
(1) When fixing the number of mesh points and increasing the basis truncation radius,
the crossover moves towards the smaller $\beta$ side.
(2) When fixing the truncation radius and increasing the number of mesh points,
the crossover moves towards the larger $\beta$ side.

\newparagraph
The solid gray line in Fig.~\ref{fig:tmp3} shows the direct space force error
for a commonly used cutoff radius of 0.9~nm.
{The splitting parameter $\beta$ should take the value
minimizing the total force error, which is the
crossover between the direct and reciprocal space errors.
Therefore, for a certain combination of the direct space cutoff, basis truncation radius and mesh size, 
we can check the intersections of the direct space error with the reciprocal space errors
using different bases to see which basis is preferred in terms of accuracy.}
{Taking $r_c = 0.9$~nm and $K_\alpha= 64$ for instance (the second plot in Fig.~\ref{fig:tmp3}),
  the Kaiser-Bessel basis is less accurate than the B-spline basis
  for $C=2$. 
  The two bases are of roughly equal accuracy for $C=3$,
  while the Kaiser-Bessel basis achieves a smaller overall error for $C=4$.
}
When decreasing the cutoff radius to $r_c = 0.6$~nm (dashed gray line)
{and fixing $K_\alpha = 64$},
the Kaiser-Bessel basis is less accurate than the B-spline basis for $C=2$,
while it is more accurate for both $C=3$ and 4.
When increasing the cutoff radius to $r_c = 1.4$~nm (dotted gray line)
{and fixing $K_\alpha = 64$},
the Kaiser-Bessel basis is less accurate for $C=2$ and $3$,
while it is as accurate as the B-spline basis for $C=4$.
{
When fixing the cutoff radius to e.g.~$r_c = 0.9$~nm,
the Kaiser-Bessel basis becomes less advantageous for a larger number of
mesh points $K_\alpha$.
  At $r_c = 0.9$~nm and $K_\alpha = 32$,
  the Kaiser-Bessel basis is more accurate for all truncations $C=2$--4. 
  At $r_c = 0.9$~nm and $K_\alpha = 96$,
  the Kaiser-Bessel basis is less accurate for $C=2$ and 3, while
  it is of the same accuracy as the B-spline basis for $C=4$.}
{At $r_c = 0.9$~nm and $K_\alpha = 128$, the Kaiser-Bessel basis is always less accurate
than the B-spline basis for all truncations $C=2$--4.}
Therefore, there is no uniform answer regarding which basis
{leads to a smaller overall error.}
It depends on the setting of other parameters.
A rule of thumb can be suggested from the results in Fig.~\ref{fig:tmp3}:
When a certain accuracy should be achieved
by a smaller direct space cutoff, less reciprocal space mesh points, and a larger truncation of basis, the Kaiser-Bessel basis may be better than the B-spline basis.


\section{Conclusion}
\label{sec:conclusion}

In this work, the Kaiser-Bessel interpolation basis {is proposed for}
the fast Ewald method.
{A reliable error estimate for the reciprocal force computation {in correlated charge systems}
is developed,
and is used to determine the optimal shape parameter of
the Kaiser-Bessel interpolation basis.}
We show that
the Kaiser-Bessel basis is more accurate in part of the working parameter space
comparing with the traditional B-spline basis that is used by the PPPM, SPME and MSME methods.
Qualitatively speaking,
this happens in the 
computation using a relatively small direct space cutoff,
a relatively small reciprocal space mesh size and a relatively large basis truncation radius.
{In practice, the better interpolation basis in terms of accuracy
can be determined, in an \emph{a priori} way,
by comparing the estimated error of the Kaiser-Bessel basis with the
estimated error of the B-spline basis~\cite{wang2010optimizing,wang2012numerical}.}

\newparagraph
{In the literature of NFFT, the Bessel and the Kaiser-Bessel basis are shown to be
always more accurate than the B-spline basis~\cite{jackson1991selection,keiner2009using},
however, Nestler~\cite{nestler2016parameter} and we demonstrate that the Bessel and the Kaiser-Bessel basis are only conditionally
{better for the fast Ewald method.}}
The reason is that the definition of ``accuracy'' for NFFT
and
that for the fast Ewald method are different.
For NFFT, it is how accurate the Fourier modes are computed, while
for the fast Ewald method it is how accurate the particle forces are computed.
Moreover, the accuracy of the fast Ewald method is also controlled by
other working parameters that do not apply in the NFFT,
i.e.~the splitting parameter and the direct space cutoff.
Therefore, the application of Kaiser-Bessel basis
in the fast Ewald method is not a trivial extension of NFFT in the particle simulation.
The optimal way of using the new basis in computing the electrostatic interaction largely relies on an accurate
error estimate, to which a great effort has been devoted by the current work.

\section*{Acknowledgment}
H.W. is supported by the National Science Foundation of China under Grants 11501039 and 91530322.
X.G. is supported by the National Science Foundation of China under Grant 91430218.
The authors gratefully acknowledge the financial support from
National High Technology Research and Development Program of China under Grant 2015AA01A304,
and National Key Research and Development Program of China under Grant 2016YFB0201200.

\appendix
\section{Error estimate for the fast Ewald method using the Kaiser-Bessel basis}
\label{app:a}
In this section, we provide key steps for deriving
the error estimate of the Kaiser-Bessel basis.
The details of the proof can be easily supplied using the
techniques in Ref.~\cite{wang2010optimizing}.
The complex exponential $e^{2\pi i mu / K}$ is approximated by
{the linear combination of the cut-off interpolation bases, i.e.}
$f(u) = \sum_{l\in I_K} g_l \myphic( u - l)$,
{where $g_l$ are coefficients that are to be determined.
  The Fourier transform of $g_l$ is denoted by $\hat g(k) = \sum_{l_\in I_K} g_l e^{-2\pi i kl/K}$,
  with the inverse transform $g_l = \frac 1K\sum_{k\in I_K}\hat g_k e^{2\pi i k/K}$.
  By definition, $f(u)$ is a convolution of $g_l$ and $\myphic$, then
  $\hat f(k) = \hat g(k)\cdot \hmyphic(k)$.
  Remembering that $f(u)$ is an approximation to the complex exponential $e^{2\pi i mu / K}$, 
  if this approximation were free of error, then $\hat f(k) = \delta_{km}$.
  Therefore, it is reasonable to assume $\hat g(k) = 0 $ for any $k\neq m$.
  Now we have $g_l = \frac1K \hat g(m) e^{2\pi i ml/K}$, and 
}
\begin{align}
  f(u) = \frac{\hat g(m)}{K} \sum_{l\in I_K} \myphic (u - l) e^{2\pi i ml/K}.
\end{align}
{We further consider the Fourier expansion of $f(u)$
\begin{align} \nonumber
  f(u)
  &=
    \sum_{k\in\mathbb Z} \hat f(k)e^{2\pi i ku/K} \\\nonumber
  &=
    \sum_{l\in\mathbb Z}\sum_{k\in I_K}  \hat f(k + lK)e^{2\pi i (k+lK) u/K} \\\nonumber
  &=
    \sum_{l\in\mathbb Z}\hat g(m)  \hmyphic(m + lK)e^{2\pi i (m+lK) u/K} \\\label{eqn:app-tmp2}
  &=
    \hat g(m)\hmyphic(m) e^{2\pi i mu/K} +
    \Big[\sum_{l\neq 0} \hat g(m) \hmyphic(m + lK) e^{2\pi ilu}\Big]\,e^{2\pi i mu/K}.
\end{align}
}
According to Eq.~\eqref{eqn:app-tmp2}, we write the approximation error of the
complex exponential (on direction $\alpha$) as
\begin{align}
  f( u_\alpha) - e^{2\pi i m_\alpha u_\alpha / K_\alpha}
  =
  e^{2\pi i m_\alpha u_\alpha / K_\alpha} E_\alpha (\vect m, \vect r),
\end{align}
where
\begin{align}
  E_\alpha(\vect m, \vect r) = \sum_l Z_{\alpha,l}(\vect m) e^{2\pi i lu_\alpha},
\end{align}
and
\begin{align}\label{eqn:app-4}
  Z_{\alpha,l}(\vect m)
  =
  \hat g(m_\alpha) \hmyphic (m_\alpha + lK_\alpha) - \delta_{l0}.
\end{align}
In the three-dimensional case, the leading order of the approximation error of the complex exponential yields
\begin{align}
  \prod_\alpha f(u_\alpha) - e^{2\pi i \vect m\cdot \vect r} =
  e^{2\pi i \vect m\cdot \vect r}
  \sum_{\alpha} E_\alpha(\vect m, \vect r).
\end{align}
The error kernel of the force computation is
\begin{align}
  \vect K(\vect r, \vect r')
  =
  \sum_{\vect m}
  \vect G(\vect m) e^{2\pi i \vect m\cdot(\vect r - \vect r')}
  \Big[
  \sum_{\alpha} E_\alpha(\vect m, \vect r)
  +
  \sum_{\alpha} E_\alpha(\vect m, -\vect r')
  \Big].
\end{align}
It can be shown, by ignoring the high frequency contributions,
\begin{align}
  \vert \vect K(\vect r, \vect r') \vert^2
  =
  2\sum_{\alpha, l} \vert (\vect G Z_{\alpha,l})^\vee(\vect r - \vect r') \vert^2.
\end{align}
The homogeneity error is therefore given by 
\begin{align}\label{eqn:app-8}
  \vert \me_\homo\vert^2 =
  2 q^2 Q^2 \sum_{\vect m} \vect G^2 (\vect m) \sum_{\alpha,l} Z^2_{\alpha,l}(\vect m).
\end{align}
{This is the Eq.~\eqref{eqn:err-esti-homo}.}
{Inserting $Z_{\alpha,l}$, saying Eq.~\eqref{eqn:app-4}, into Eq.~\eqref{eqn:app-8}, we have
\begin{align}
  \vert \me_\homo\vert^2 =
  2 q^2 Q^2 \sum_{\vect m} \vect G^2 (\vect m)
  \sum_\alpha
  \Big\{
  \hat g^2(m_\alpha) \sum_{l}  \hmyphic^2 (m_\alpha + lK_\alpha)
  -
  2 \hat g(m_\alpha) \hmyphic (m_\alpha)
  +
  1
  \Big\}.
\end{align}
When $\hat g(m_\alpha)$ takes
\begin{align}\label{eqn:app-tmp11}
  \hat g(m_\alpha) =
  \frac{\hmyphic(m_\alpha)}{\sum_l \hmyphic^2(m_\alpha + lK_\alpha) },
\end{align}
the homogeneity error is minimized to
\begin{align}
  \vert \me_\homo\vert^2 =
  2 q^2 Q^2 \sum_{\vect m} \vect G^2 (\vect m)
  \sum_\alpha
  \frac
  {\sum_{l\neq 0} \hmyphic^2(m_\alpha + lK_\alpha) }  
  {\sum_l \hmyphic^2(m_\alpha + lK_\alpha) }.
\end{align}
Inserting Eq.~\eqref{eqn:app-tmp11} into Eq.~\eqref{eqn:app-4}, we have
\begin{align}
  Z_{\alpha,l}(\vect m)
  =
  \frac{\hmyphic(m_\alpha)\hmyphic (m_\alpha + lK_\alpha)}
  {\sum_l \hmyphic^2(m_\alpha + lK_\alpha) } - \delta _{l0},
\end{align}
which was used in in the error estimate Eq.~\eqref{eqn:app-8}, and gives Eq.~\eqref{eqn:err-esti-homo-z}.
}


\begin{thebibliography}{10}

\bibitem{cisneros2013classical}
G~Andr\'es Cisneros, Mikko Karttunen, Pengyu Ren, and Celeste Sagui.
\newblock Classical electrostatics for biomolecular simulations.
\newblock {\em Chemical reviews}, 114(1):779--814, 2014.

\bibitem{onsager1936electric}
Lars Onsager.
\newblock Electric moments of molecules in liquids.
\newblock {\em Journal of the American Chemical Society}, 58(8):1486--1493,
  1936.

\bibitem{vangunsteren1978inclusion}
W.F. van Gunsteren, H.J.C. Berendsen, and J.A.C. Rullmann.
\newblock Inclusion of reaction fields in molecular dynamics. application to
  liquid water.
\newblock {\em Faraday Discuss. Chem. Soc.}, 66:58--70, 1978.

\bibitem{wolf1999exact}
D.~Wolf, P.~Keblinski, SR~Phillpot, and J.~Eggebrecht.
\newblock Exact method for the simulation of coulombic systems by spherically
  truncated, pairwise $r^{-1}$ summation.
\newblock {\em The Journal of chemical physics}, 110:8254, 1999.

\bibitem{wu2005isotropic}
Xiongwu Wu and Bernard~R Brooks.
\newblock Isotropic periodic sum: A method for the calculation of long-range
  interactions.
\newblock {\em The Journal of chemical physics}, 122(4):044107, 2005.

\bibitem{fukuda2011molecular}
I.~Fukuda, Y.~Yonezawa, and H.~Nakamura.
\newblock Molecular dynamics scheme for precise estimation of electrostatic
  interaction via zero-dipole summation principle.
\newblock {\em The Journal of chemical physics}, 134:164107, 2011.

\bibitem{fukuda2013zero}
I.~Fukuda.
\newblock Zero-multipole summation method for efficiently estimating
  electrostatic interactions in molecular system.
\newblock {\em The Journal of Chemical Physics}, 139:174107, 2013.

\bibitem{wang2016critical}
Han Wang, Haruki Nakamura, and Ikuo Fukuda.
\newblock A critical appraisal of the zero-multipole method: Structural,
  thermodynamic, dielectric, and dynamical properties of a water system.
\newblock {\em The Journal of chemical physics}, 144(11):114503, 2016.

\bibitem{ewald1921die}
P.~P. Ewald.
\newblock Die berechnung optischer und elektrostatischer gitterpotentiale.
\newblock {\em Ann. Phys.}, 369(3):253--287, 1921.

\bibitem{darden1993pme}
T.~Darden, D.~York, and L.~Pedersen.
\newblock Particle mesh ewald: An n· log (n) method for ewald sums in large
  systems.
\newblock {\em The Journal of Chemical Physics}, 98:10089, 1993.

\bibitem{essmann1995spm}
U.~Essmann, L.~Perera, M.L. Berkowitz, T.~Darden, H.~Lee, and L.G. Pedersen.
\newblock A smooth particle mesh ewald method.
\newblock {\em The Journal of Chemical Physics}, 103(19):8577, 1995.

\bibitem{deserno1998mue1}
M.~Deserno and C.~Holm.
\newblock How to mesh up ewald sums. i. a theoretical and numerical comparison
  of various particle mesh routines.
\newblock {\em The Journal of Chemical Physics}, 109:7678, 1998.

\bibitem{cerutti2009multi}
D.S. Cerutti and D.A. Case.
\newblock Multi-level ewald: a hybrid multigrid/fast fourier transform approach
  to the electrostatic particle-mesh problem.
\newblock {\em Journal of chemical theory and computation}, 6(2):443--458,
  2009.

\bibitem{wang2016multiple}
Han Wang, Xingyu Gao, and Jun Fang.
\newblock Multiple staggered mesh ewald: Boosting the accuracy of the smooth
  particle mesh ewald method.
\newblock {\em arXiv preprint arXiv:1607.04008}, 2016.

\bibitem{arnold2013comparison}
A.~Arnold, F.~Fahrenberger, C.~Holm, O.~Lenz, M.~Bolten, H.~Dachsel, R.~Halver,
  I.~Kabadshow, F.~G{\"a}hler, F.~Heber, J.~Iseringhausen, M.~Hofmann,
  M.~Pippig, D.~Potts, and G.~Sutmann.
\newblock Comparison of scalable fast methods for long-range interactions.
\newblock {\em Physical Review E}, 88(6):063308, 2013.

\bibitem{ballenegger2012convert}
V.~Ballenegger, J.J. Cerd{\`a}, and C.~Holm.
\newblock How to convert spme to p3m: influence functions and error estimates.
\newblock {\em Journal of Chemical Theory and Computation}, 8:936--947, 2012.

\bibitem{dutt1993fast}
Alok Dutt and Vladimir Rokhlin.
\newblock Fast fourier transforms for nonequispaced data.
\newblock {\em SIAM Journal on Scientific computing}, 14(6):1368--1393, 1993.

\bibitem{steidl1998note}
Gabriele Steidl.
\newblock A note on fast fourier transforms for nonequispaced grids.
\newblock {\em Advances in computational mathematics}, 9(3-4):337--352, 1998.

\bibitem{ware1998fast}
Antony~F Ware.
\newblock Fast approximate fourier transforms for irregularly spaced data.
\newblock {\em SIAM review}, 40(4):838--856, 1998.

\bibitem{fourmont2003non}
Karsten Fourmont.
\newblock Non-equispaced fast fourier transforms with applications to
  tomography.
\newblock {\em Journal of Fourier Analysis and Applications}, 9(5):431--450,
  2003.

\bibitem{fessler2003nonuniform}
Jeffrey~A Fessler and Bradley~P Sutton.
\newblock Nonuniform fast fourier transforms using min-max interpolation.
\newblock {\em Signal Processing, IEEE Transactions on}, 51(2):560--574, 2003.

\bibitem{greengard2004accelerating}
Leslie Greengard and June-Yub Lee.
\newblock Accelerating the nonuniform fast fourier transform.
\newblock {\em SIAM review}, 46(3):443--454, 2004.

\bibitem{keiner2009using}
Jens Keiner, Stefan Kunis, and Daniel Potts.
\newblock Using nfft 3---a software library for various nonequispaced fast
  fourier transforms.
\newblock {\em ACM Transactions on Mathematical Software (TOMS)}, 36(4):19,
  2009.

\bibitem{pippig2013parallel}
Michael Pippig and Daniel Potts.
\newblock Parallel three-dimensional nonequispaced fast fourier transforms and
  their application to particle simulation.
\newblock {\em SIAM Journal on Scientific Computing}, 35(4):C411--C437, 2013.

\bibitem{hedman2006ewald}
F.~Hedman and A.~Laaksonen.
\newblock Ewald summation based on nonuniform fast fourier transform.
\newblock {\em Chemical physics letters}, 425(1-3):142--147, 2006.

\bibitem{nestler2015fast}
Franziska Nestler, Michael Pippig, and Daniel Potts.
\newblock Fast ewald summation based on nfft with mixed periodicity.
\newblock {\em Journal of Computational Physics}, 285:280--315, 2015.

\bibitem{potts2003fast}
Daniel Potts and Gabriele Steidl.
\newblock Fast summation at nonequispaced knots by nfft.
\newblock {\em SIAM Journal on Scientific Computing}, 24(6):2013--2037, 2003.

\bibitem{jackson1991selection}
John~I Jackson, Craig~H Meyer, Dwight~G Nishimura, and Albert Macovski.
\newblock Selection of a convolution function for fourier inversion using
  gridding [computerised tomography application].
\newblock {\em Medical Imaging, IEEE Transactions on}, 10(3):473--478, 1991.

\bibitem{nestler2016parameter}
Franziska Nestler.
\newblock Parameter tuning for the nfft based fast ewald summation.
\newblock {\em Frontier in Physics}, 4, 2016.

\bibitem{deserno1998mue2}
M.~Deserno and C.~Holm.
\newblock How to mesh up ewald sums. ii. an accurate error estimate for the
  particle--particle--particle-mesh algorithm.
\newblock {\em The Journal of Chemical Physics}, 109:7694, 1998.

\bibitem{jorgensen1983comparison}
William~L Jorgensen, Jayaraman Chandrasekhar, Jeffry~D Madura, Roger~W Impey,
  and Michael~L Klein.
\newblock Comparison of simple potential functions for simulating liquid water.
\newblock {\em The Journal of chemical physics}, 79(2):926--935, 1983.

\bibitem{ballenegger2009simulations}
V~Ballenegger, A~Arnold, and JJ~Cerda.
\newblock Simulations of non-neutral slab systems with long-range electrostatic
  interactions in two-dimensional periodic boundary conditions.
\newblock {\em The Journal of chemical physics}, 131(9):094107, 2009.

\bibitem{de1980simulation}
Simon~W de~Leeuw, John~William Perram, and Edgar~Roderick Smith.
\newblock Simulation of electrostatic systems in periodic boundary conditions.
  i. lattice sums and dielectric constants.
\newblock In {\em Proceedings of the Royal Society of London A: Mathematical,
  Physical and Engineering Sciences}, volume 373, pages 27--56. The Royal
  Society, 1980.

\bibitem{de1980simulation2}
SW~De~Leeuw, JW~Perram, and ER~Smith.
\newblock Simulation of electrostatic systems in periodic boundary conditions.
  ii. equivalence of boundary conditions.
\newblock In {\em Proceedings of the Royal Society of London A: Mathematical,
  Physical and Engineering Sciences}, volume 373, pages 57--66. The Royal
  Society, 1980.

\bibitem{frenkel2001understanding}
D.~Frenkel and B.~Smit.
\newblock {\em Understanding molecular simulation}.
\newblock Academic Press, Inc. Orlando, Fl, USA, 2010.

\bibitem{perram1988asc}
J.~Perram, H.~Petersen, and S.~De~Leeuw.
\newblock An algorithm for the simulation of condensed matter which grows as
  the 3/2 power of the number of particles.
\newblock {\em Molecular Physics}, 65(4):875--893, 1988.

\bibitem{kolafa1992cutoff}
J.~Kolafa and J.W. Perram.
\newblock Cutoff errors in the ewald summation formulae for point charge
  systems.
\newblock {\em Molecular Simulation}, 9(5):351--368, 1992.

\bibitem{wang2012error}
H.~Wang, C.~Sch{\"u}tte, and P.~Zhang.
\newblock Error estimate of short-range force calculation in inhomogeneous
  molecular systems.
\newblock {\em Physical Review E}, 86(2):026704, 2012.

\bibitem{wang2012numerical}
H.~Wang, P.~Zhang, and C.~Sch{\"u}tte.
\newblock On the numerical accuracy of ewald, smooth particle mesh ewald, and
  staggered mesh ewald methods for correlated molecular systems.
\newblock {\em Journal of Chemical Theory and Computation}, 8(9):3243--3256,
  2012.

\bibitem{gao2016sampling}
Xingyu Gao, Jun Fang, and Han Wang.
\newblock Sampling the isothermal-isobaric ensemble by langevin dynamics.
\newblock {\em The Journal of chemical physics}, 144(12):124113, 2016.

\bibitem{wang2010optimizing}
H.~Wang, F.~Dommert, and C.~Holm.
\newblock Optimizing working parameters of the smooth particle mesh ewald
  algorithm in terms of accuracy and efficiency.
\newblock {\em The Journal of chemical physics}, 133:034117, 2010.

\end{thebibliography}

\end{document}